\documentclass{eas}
\usepackage{graphicx}
\begin{document}

\TitreGlobal{Mass Profiles and Shapes of Cosmological Structures}

\title{Lost and found dark matter in elliptical galaxies}
\author{Stoehr, F.}\address{IAP, 98 bis Bd Arago, F-75014 Paris, FRANCE}
\author{Mamon, G. A.$^1$}
\author{Dekel, A.}\address{Hebrew University, Jerusalem, ISRAEL}
\author{Cox, T. J.}\address{Harvard-Smithsonian 
CfA, 60 Garden St, Cambridge MA 02138, USA}
\runningtitle{Dark matter in ellipticals}
\setcounter{page}{135}
\index{Stoehr, F.}
\index{Mamon, G. A.}
\index{Dekel, A.}
\index{Cox, T. J.}

%
\begin{abstract}
The kinematical properties of elliptical galaxies formed during the mergers
of equal mass, stars+gas+dark matter spiral galaxies are compared to the
observed low velocity 
dispersions found for planetary nebulae on the outskirts of ellipticals,
which have been interpreted as pointing to a lack of dark matter in
ellipticals 
(which poses a problem for the standard model of galaxy formation).
We find that the velocity dispersion profiles of the stars in the
simulated ellipticals
match well the observed ones.
The low outer stellar velocity dispersions are mainly caused by the radial
orbits of the outermost stars, which, for a given binding energy must have
low angular momentum to reach their large radial distances, usually driven  
out along tidal tails.
\end{abstract}
\maketitle
%
\section{Introduction}
There is a wide consensus that spiral galaxies must be embedded within dark
matter halos, as there have been no other good explanations of the observed 
flat rotation curves of spiral galaxies, unless one resorts to modifying
physics (e.g. MOND, see McGaugh in these proceedings).
Moreover, dissipationless cosmological $N$-body simulations lead to
structures, whose halos represent most spiral galaxies
(Hayashi et el. 2005; Stoehr 2005).

If elliptical galaxies originate from major mergers of spiral galaxies (Toomre
1977; Mamon 1992; Baugh, Cole \& Frenk 1995; Springel et al. 2001a), then
they too should possess dark matter halos.
Using planetary nebulae (PNe) as tracers of the dark matter at large radii,
Romanowsky et al. (2003) found low velocity dispersions for their outermost
PNe, which after some simple Jeans modeling and more sophisticated orbit
modeling led them to conclude to a dearth of dark matter in ordinary
elliptical galaxies. This result is not expected in the standard model of
structure and galaxy formation.
This has led us (Dekel et al. 2005) to analyze the final outputs of $N$-body
simulations of spiral galaxies merging into ellipticals.

\section{Merger simulations}

\begin{figure}[ht]
\centering
\includegraphics[width=10.2cm]{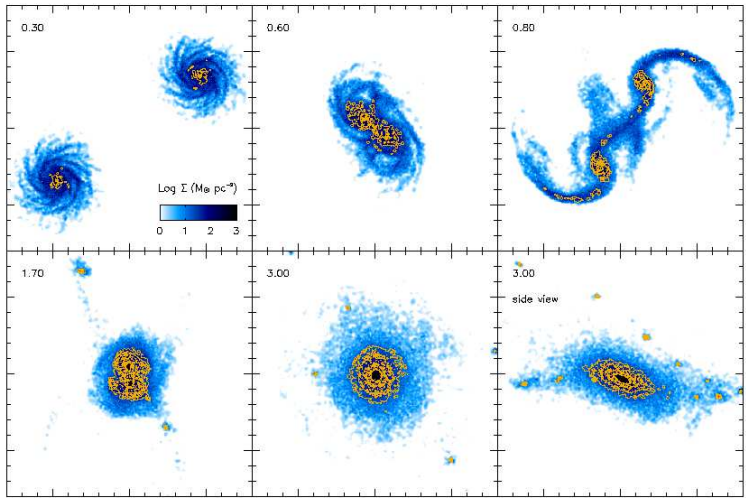}
\caption{Snapshots of two equal mass spiral galaxies merging into
a single elliptical galaxy. From Dekel et al. (2005, see also Cox 2004, Cox
et al. 2004).}
\label{snapshots}
\end{figure}

The merger simulations we have analyzed were run by Cox (2004, see also Cox et
al. 2004). In these simulations, the initial spiral galaxies had an
exponential spherical bulge and
a thin exponential 
disk, as well as a thin gaseous exponential disk and a spherical NFW (Navarro et
al. 1996) dark matter halo.
The particles were advanced with the {\sf GADGET} TREE-SPH code (Springel et
al. 2001b) until
2--3 Gyr after the final merger.
The galaxies were thrown at one another on a variety of near parabolic orbits
(see Fig.~\ref{snapshots}).

\section{2D diagnostics}

\begin{figure}[ht]
\centering
\includegraphics[width=5.5cm]{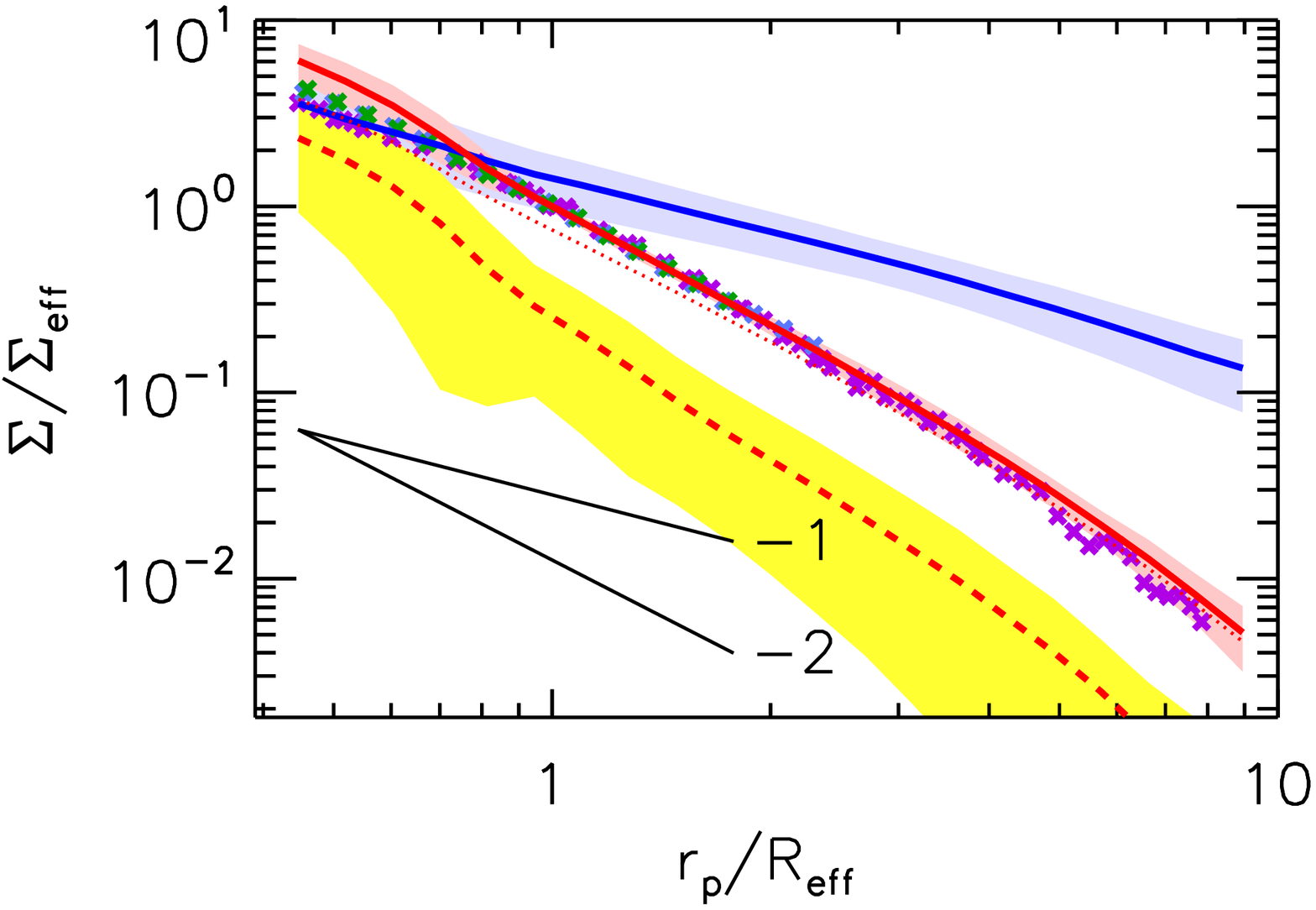}
\includegraphics[width=6.7cm]{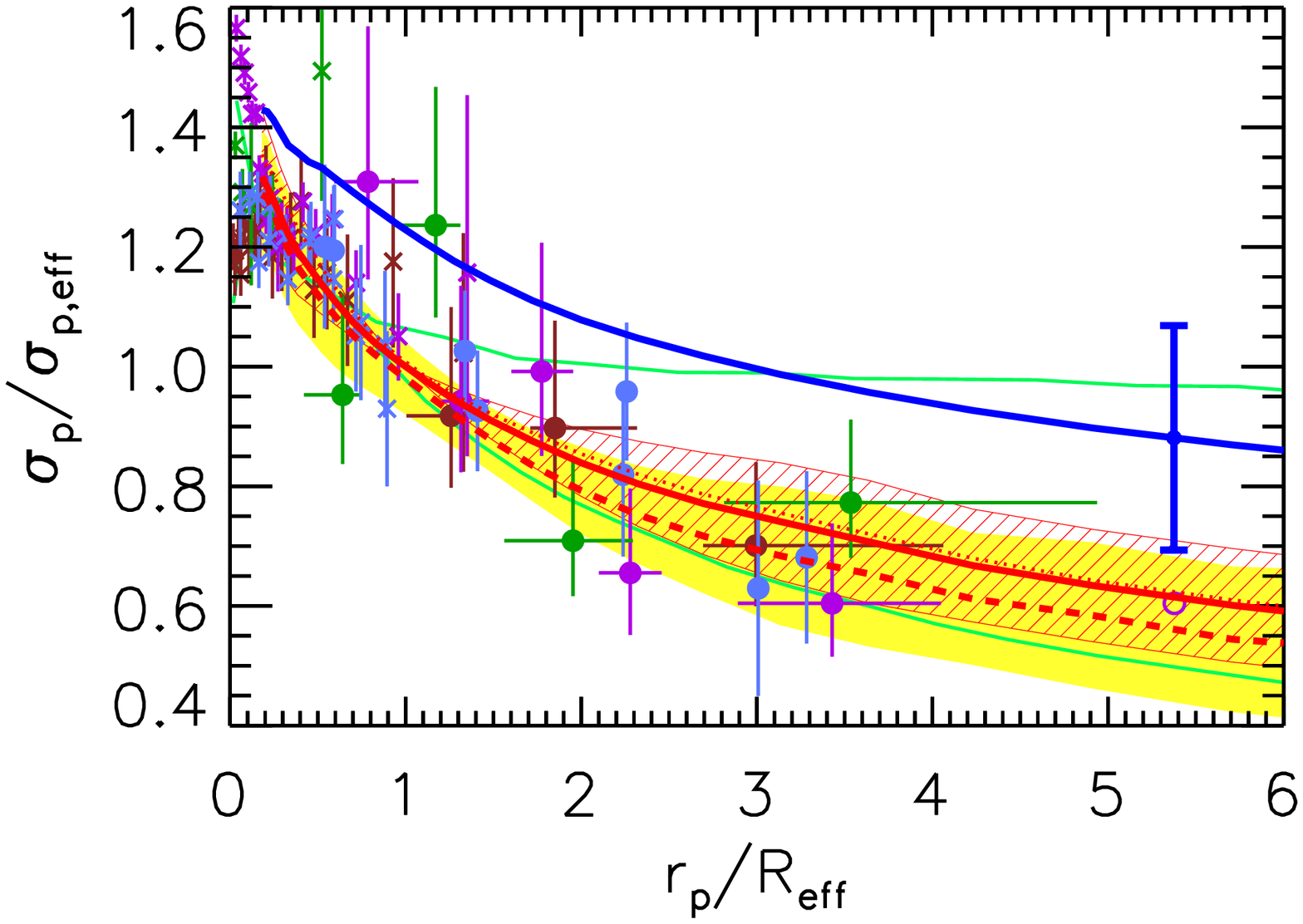}
\caption{\emph{Left}:
Surface density (\emph{curves}) and brightness (\emph{symbols}) 
profiles of
simulated (\emph{curves}) and observed (\emph{symbols}: NGC~3379 and NGC~821)
elliptical galaxies.  
\emph{Right}: Line-of-sight velocity dispersion profiles. 
The \emph{lower} and \emph{upper thin (green) curves}
represent the predictions of Romanowsky et al. (2003), respectively without
and with dark matter. From Dekel et al. (2005).
\emph{Both plots}: dark matter are the \emph{upper solid (blue) curves},
stars are the \emph{lower curves} (old, young and `all' are \emph{dotted},
\emph{dashed} 
and \emph{solid}).
The \emph{curves} are averages over 60 profiles (10 simulations, 2 timesteps
and 3 viewing angles), 
and the \emph{shaded regions} indicate the $\pm1\,\sigma$ spread.
The $x-$ and $y-$ axes are normalized to the values at the half-projected
light (mass) radius, i.e. effective radius, $R_{\rm eff}$.
}
\label{2dplots}
\end{figure}

The left hand plot of
Figure~\ref{2dplots} shows the surface density of the different components
of the simulated galaxies, and the surface brightness profiles of galaxies
NGC~3379 and NGC~821.
The match is excellent, except perhaps in the very inner regions, where the
simulations predict too much mass. 

More important, the right-hand plot of Figure~\ref{2dplots} shows that 
the simulated and observed velocity dispersion
profiles are very similar.
In other words, \emph{our simulations, which include normal amounts of dark
matter, 
reproduce very well the observed velocity dispersion profiles}. Hence, we
conclude that \emph{the 
low outer velocity dispersions do not imply a lack of dark
matter in elliptical galaxies}.


\section{3D diagnostics}

What causes this discrepancy between the \emph{kinematic}
modeling of 
Romanowsky et al., which concludes to no or little dark matter in ellipticals,
and the \emph{dynamical} modeling, which concludes to a normal content of
dark matter in ellipticals?
\begin{figure}[ht]
\centering
\includegraphics[width=6cm]{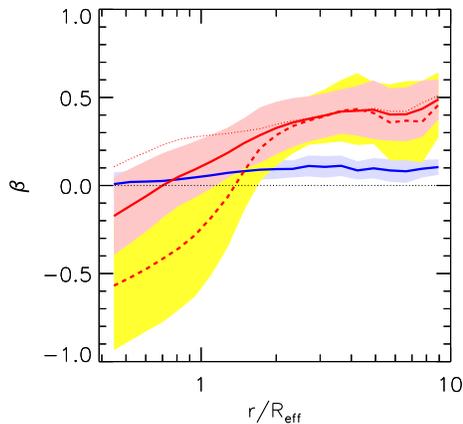}
\caption{Velocity anisotropy versus radius (from Dekel et al. 2005).
Dark matter is the \emph{nearly horizontal curve}, while the other curves are
for the stars (\emph{dotted}, \emph{dashed} and \emph{solid} for old, young
and `all' stars, respectively).
}
\label{beta}
\end{figure}
Figure~\ref{beta} shows that for $R > 2\,R_{\rm eff}$, while the
dark matter particles travel along very slightly radial orbits, as
found in cosmological simulations (see Mamon \& {\L}okas 2005, and references
therein), \emph{the outer stellar particles travel along very elongated
orbits}. 

Why do the stars at a few $R_{\rm eff}$ travel on much more elongated orbits
than do the dark matter particles at the same distances from the elliptical
galaxy center?
Since the stars in the initial
spiral galaxies lie at small radii (while the dark matter particles lie in a
wide range of radii), they will usually 
end up at small radii in the
merging pair, since the most bound particles usually remain the most bound 
(Barnes 1992). 
So the outermost stars must have traveled along elongated orbits to reach
their large present radial distances: i.e., not only they have low binding
energies, but they must have low angular momentum given their binding energy.
A more detailed analysis indicates that these stars are driven outwards on
nearly radial orbits in the tidal tails (see Fig.~\ref{snapshots})
that are formed after the first
passage of the two spirals.

In a companion contribution (Mamon, in these proceedings), we analyze in
further detail what quantity of dark matter can be derived from kinematical
modeling and what we can learn from the detailed structure and internal
kinematics of the simulated merger remnants.

\end{document}